\documentclass[reprint,superscriptaddress,amsmath,amssymb,amsfonts,aps,pra]{revtex4-1}

\usepackage{graphicx}
\usepackage{dcolumn}
\usepackage{bm}
\usepackage{hyperref}
\usepackage{subfigure}
\usepackage{qcircuit}
\usepackage{bbold}
\usepackage{xcolor}
\usepackage{soul}

\newcommand{\bra}[1]{\mbox{$\langle #1|$}}
\newcommand{\ket}[1]{\mbox{$|#1\rangle$}}

\begin{document}

\title{Certified variational quantum algorithms for eigenstate preparation}

\author{Andrey Kardashin}
\email{andrey.kardashin@skoltech.ru}
\author{Alexey Uvarov}
\author{Dmitry Yudin}
\author{Jacob Biamonte}
\affiliation{Deep Quantum Laboratory, Skolkovo Institute of Science and Technology, Moscow 121205, Russia}

\date{\today}

\begin{abstract}
Solutions to many-body problem instances often involve an intractable number of degrees of freedom and admit no known approximations in general form. In practice, representing quantum-mechanical states of a given Hamiltonian using available numerical methods, in particular those based on variational Monte Carlo simulations, become exponentially more challenging with increasing system size. Recently quantum algorithms implemented as variational models, have been proposed to accelerate such simulations. The variational ansatz states are characterized by a polynomial number of parameters devised in a way to minimize the expectation value of a given Hamiltonian, which is emulated by local measurements. In this study, we develop a means to certify the termination of variational algorithms. We demonstrate our approach by applying it to three models: the transverse field Ising model, the model of one-dimensional spinless fermions with competing interactions, and the Schwinger model of quantum electrodynamics. By means of comparison, we observe that our approach shows better performance near critical points in these models. We hence take a further step to improve the applicability and to certify the results of variational quantum simulators. 
\end{abstract}

\maketitle

\section{Introduction}
Experimental advances have fostered the development of midsized quantum simulators---realizing prototypes of ideas dating back to celebrated proposals by Feynman and others \cite{Feynman1982,Lloyd1996,Buluta2009,Brown2010,Hauke2012,Schaetz2013,Georgescu2014}. Indeed, controllable quantum simulators emulate classes of Hamiltonians---mimicking Hamiltonian properties to replace traditional numerical methods \cite{Parsons2016,Bremner2016,Gao2017,Vega2018,Gluza2020}. The difficultly of numerical simulations of interacting quantum systems has resulted in advanced numerical methods, including variational quantum Monte Carlo methods \cite{Foulkes2001} as well as different realizations of the renormalization group routine \cite{Bulla2008}, being computationally intractable.  In limiting cases, these methods suffer from the exponential slowdown (and/or exponential memory overhead) with the size of a system. 

Multiqubit quantum circuits can implement the so-called variational model of quantum computation \cite{Peruzzo2014, Clean2016, Akshay_2020}, which extends certain methods of machine learning \cite{Lecun2015, Biamonte2017}. In the variational quantum circuits approach, one relies on an iterative control loop. A quantum state is prepared and measured: The measurement outcome(s) are used to prepare increasingly more optimal states with respect to minimization of a given objective function (given as a Hamiltonian). Variational algorithms emerged as a practically viable application of quantum computers with several dozen qubits and short decoherence times which would otherwise preclude the use of more traditional quantum algorithms \cite{yung_transistor_2014,Peruzzo2014,Clean2016, Akshay_2020}. The results of measurement are used in a classical optimization routine to update the prepared state so as to minimize an externally calculated objective function. The process is iterated and the states are prepared by varying over a family of low-depth circuits. 

Although experimental realizations of variational algorithms \cite{Preskill2018,Moll2018} were reported in recent years \cite{Kokail2019,LaRose2019}, theoretical estimates of their efficiency \cite{Akshay_2020} are largely lacking. A particular example, the variational quantum eigensolver (VQE), prepares a family of states characterized by a polynomial number of parameters and minimizes the expectation value of a given Hamiltonian within this family \cite{Peruzzo2014,PhysRevX.6.031007,kandala_hardware-efficient_2017}. The key idea of VQE is based on decomposing the Hamiltonian into a sum of Pauli strings, i.e., tensor products of Pauli matrices, provided that each Pauli string can be measured separately on the quantum device. VQE can be applied to find ground states of small molecules and interacting spin systems \cite{PhysRevX.6.031007,PhysRevA.95.020501}.  Scaling of such an approach could access simulations that  are not possible to evaluate explicitly using traditional numerical methods, for example, owing to the lack of memory or computational resources.

The performance of VQE crucially depends on the choice of the ansatz state. Typically, a common approach is to represent a rather cumbersome quantum state in terms of a variational state and estimate approximation quality, i.e., to explore how close the obtained solution is to the ground state of a given Hamiltonian. Knowing an exact solution drastically simplifies the analysis; otherwise, the proximity to the global minimum cannot be guaranteed. Generally, minimization of Hamiltonians is QMA-hard, whereas its restriction to Ising spins is NP-hard. Lately, a way to estimate the quality of the solution by measuring the variance of the energy has been proposed in Ref.~\cite{Kokail2019}. 

In the scope of this paper, we propose an alternative approach by simulating the Hamiltonian evolution. We clearly demonstrate that in this scenario the number of measurements can be dramatically reduced. We consider the two competing criteria as optimization problems on their own, aside from the VQE problem. We compare the convergence of the two algorithms and clarify the limits of applicability of our method, with a special focus on connection between computational complexity of Hamiltonians and the properties of their eigenstates that are parametrized in terms of the hardware efficient ansatz, that is specifically tailored to the available interactions in a quantum processor.

\begin{figure*}
    \includegraphics[width=.75\textwidth]{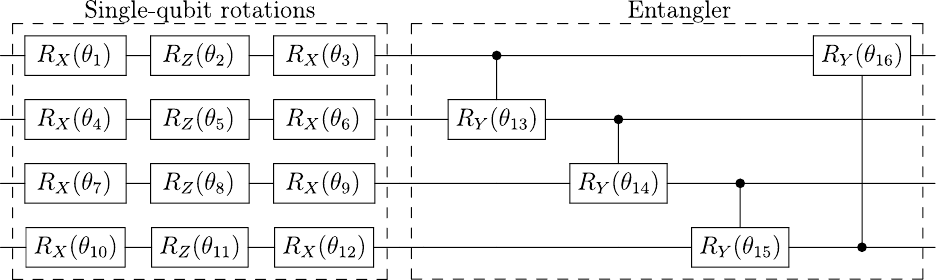}
    \caption{Hardware efficient ansatz for four qubits with $R_{\sigma}(\theta) = e^{-i\theta\sigma}$, $\sigma\in \{\mathbb{1}, X, Y, Z\}$. The entire layer can be repeated several times as needed.}
    \label{fig:hardware_efficient_ansatz}
\end{figure*}

\section{Variational eigenvector search}
The problem we solve is somewhat complementary to VQE. Given a Hamiltonian, defined by its Hermitian matrix, find an eigenvector of this Hamiltonian. The apparent simplicity of determining the eigenvectors of a given matrix nevertheless obscures its computational complexity. It can be done either by means of exact diagonalization, e.g., leveraging Lanczos algorithm, or variational-ansatz-based simulations, both being computationally demanding \cite{Golub1996}.

Consider the problem of finding an eigenvector of a Hermitian matrix using a variational quantum algorithm approach. VQE, at its core, relies on preparing an ansatz state $\vert\psi(\boldsymbol{\theta})\rangle$ by applying an adjustable sequence of quantum gates $U(\bm{\theta})$ to the quantum register $\ket{\boldsymbol{0}} \equiv \ket{0}^{\otimes n}$ of $n$ qubits and sampling the expectation value of a given matrix $\mathcal{H}$ relative to this state.  This is followed by a classical optimizer to minimize the energy, $\langle\psi(\boldsymbol{\theta})\vert \mathcal{H}\vert\psi(\boldsymbol{\theta})\rangle$. The circuit is parametrized by $\boldsymbol{\theta} \in [0, 2\pi)^{\times p}$ with $p$ being the number of parameters. Assume that within VQE our best guess is $\ket{\psi}\equiv\ket{\psi(\bm{\theta})}$. 
In Ref.~\cite{Kokail2019}, to quantify the accuracy of the variational solution $\ket{\psi}$ it was proposed to employ the mean squared deviation, $\delta=\langle\mathcal{H}^2\rangle-\langle\mathcal{H}\rangle^2$ (note that we make use of notation $\langle A\rangle=\bra{\psi}A\ket{\psi}$ below). 
In fact, let the eigenenergy $\lambda_0$ be the closest to the initial trial $\mathcal{E}$; then the energy error is upper bound by $\delta$,
\begin{equation}
    \vert\mathcal{E}-\lambda_0\vert\leqslant\sqrt{\delta}.
\end{equation}
The vector $\ket{\psi}$ is an eigenvector of the Hermitian $\mathcal{H}$ if and only if the mean squared variance $\delta$ is zero. Alternatively, a unitary matrix $\mathcal{Q}=e^{-i\mathcal{H}t}$ possesses eigenvalues lying on the unit circle, so that $\ket{\psi}$ is an eigenvector of $\mathcal{H}$ as long as $\vert\bra{\psi}\mathcal{Q}\ket{\psi}\vert = 1$.

In numerical simulations, we choose the unitary $U(\boldsymbol{\theta})$ to be parametrized in terms of three-layered hardware-efficient ansatz as depicted in Fig.~\ref{fig:hardware_efficient_ansatz}. By construction, the hardware efficient ansatz --- first introduced in Ref.~\cite{kandala_hardware-efficient_2017} --- consists of an array of universal one-qubit gates and an entangling block. The universal one-qubit gates are represented in the $X$-$Z$ decomposition while the entangling block is composed of subsequent controlled $Y$ rotations. The $m$-layered $n$-qubit ansatz would have $4mn$ parameters for $n > 2$. In this study, we use a four-qubit ansatz with $m=3$ layers, and therefore, 48 free parameters. The parameters $\boldsymbol{\theta}$ are updated by means of the Broyden-Fletcher-Goldfarb-Shanno (BFGS) algorithm \cite{NoceWrig06}, which is a gradient-based method that uses an approximation of the Hessian matrix.

\section{Model systems}
In the following, we address the convergence properties of physically relevant systems. We consider a one-dimensional quantum Ising chain of $n$ spins, which corresponds to the number of qubits,
\begin{equation}\label{eq:tfim}
    \mathcal{H}_{\text{TFIM}} = \mathcal{J}\sum_{j=1}^n \left( Z_j Z_{j+1} + h X_j \right),
\end{equation}
in the presence of transverse magnetic field $h\mathcal{J}$ with $\mathcal{J}$ specifying the strength of exchange interaction \cite{Dutta2015}. Note that $(\mathbb{1}_j,X_j,Y_j,Z_j)$ stands for the vector of Pauli matrices at the $j$th site equipped with a $2\times2$ unity matrix. In the thermodynamic limit, $n\rightarrow\infty$, the system undergoes the phase transition from a collinearly ordered to a disordered phase at $h=1$, which will be discussed in the follow-up analysis. Quite interestingly, recent analysis based on neural networks machinery in the form of single \cite{Berezutskii2020} and multilayer perceptron \cite{Arai2018} demonstrated its efficiency in studying phase transition for the model of Eq.~(\ref{eq:tfim}). 

Likewise, we examine our method to find an eigenstate of the massive Schwinger Hamiltonian,
\begin{equation}\label{eq:sch}
    \mathcal{H}_{\text{Sch}} = \sum_{j=1}^n \Big[ \sigma_{j}^+ \sigma_{j+1}^- + \sigma_{j}^- \sigma_{j+1}^+ + \frac{m_c}{2} (-1)^j Z_j + L_j^2 \Big],
\end{equation}
provided that $L_j = -\frac{1}{2}\sum_{i=1}^j \left[ Z_i + (-1)^i \mathbb{1}_i\right]$ and $\sigma^\pm_j = \left(X_j \pm i Y_j \right)/2$. The model (\ref{eq:sch}) has remained in the focus of research activity as it allows one to capture intriguing properties of quantum chromodynamics. In a nutshell, the Schwinger model represents quantum electrodynamics in two-dimensional space-time \cite{Hamer1997} and can be addressed in a seemingly related approach of matrix product states; see, e.g., \cite{Buyens2016}. In our simulations, we put $m_c = -0.7$ that corresponds to criticality of this model.

Finally we consider a system of one-dimensional spinless fermions with competing interactions,
\begin{equation}\label{eq:el}
    \mathcal{H}_{\text{el}} = -t\sum_{\langle i,j\rangle} c^\dagger_i c_j + U_1\sum_{j=1}^nn_j n_{j+1} + U_2\sum_{j=1}^nn_j n_{j+2},
\end{equation}
where $n_j=c_j^\dagger c_j$ is the number of electrons at the $j$th site and summation over nearest neighbors $\langle\ldots\rangle$ is implied. In this model, $t$ is the hopping energy, while $U_1$ and $U_2$ stand for matrix elements of Coulomb repulsion between electrons residing on two neighboring and next-neighboring sites respectively. The model (\ref{eq:el}) represents a versatile still rather simple playground to study effects of frustration in interacting systems \cite{Zhuravlev2000,Karrasch2012,Hohenadler2012,Uvarov2020}. With the fixed ratio $U_1/U_2 = 2$, this model is expected to exhibit a metallic behavior \cite{Zhuravlev2000}. In contrast to the models of Eqs.~(\ref{eq:tfim}) and (\ref{eq:sch}), the Hamiltonian of interacting electrons (\ref{eq:el}) is written in terms of second-quantized fermionic annihilation ($c_j$) and creation ($c_j^\dagger$) operators, which requires spin-fermion mapping to be implemented. We utilize the Jordan-Wigner transformation to represent these operators as
\begin{equation*}
    c_j^\dagger = \left( \bigotimes_{k=1}^{j-1} Z_k \right) \otimes \sigma^+_j, 
    \quad
    c_j = \left( \bigotimes_{k=1}^{j-1} Z_k \right) \otimes \sigma^-_j.
\end{equation*}

\section{Cost function}
To make a direct comparison with the results of the previous studies \cite{Kokail2019} and discuss the range of applicability of our method, we consider the performance of two cost functions determined by
\begin{eqnarray}\label{eq:cost_var}
    F_H(\psi) = \langle\mathcal{H}^2 \rangle - \langle \mathcal{H}\rangle^2, \\ \label{eq:cost_uni}
    F_Q(\psi) = 1 - \big|\langle\mathcal{Q}\rangle \big|^2,
\end{eqnarray}
respectively (see Appendix~\ref{app:appa} for more details). Notably, both functions return zero if and only if $\ket{\psi}$ is an eigenstate of the matrix $\mathcal{H}$. To control the efficiency of both methods, we apply the \textit{gain} characteristic as a quantitative measure \cite{wilson2019optimizing},
\begin{equation} \label{eq:gain}
    \mathcal{G} = \mathbb{E}_F \left[ \frac{F_\mathrm{conv} - F_\mathrm{init}}{F_\mathrm{opt} - F_\mathrm{init}} \right],
\end{equation}
representing the mean variance of the cost function over all instances and written in terms of the value of the objective function at the start of optimization ($F_\mathrm{init}$) and at the end of convergence ($F_\mathrm{conv}$), as well as the optimal value of  ($F_\mathrm{opt}$), i.e., global minimum or maximum. Likewise, we elaborate on gain of the overlap $\mathcal{O}$ between the variational state $\ket{\psi}$ and an exact eigenstate $\ket{\varphi}$ of a target Hamiltonian, i.e.,
\begin{equation}
    \label{eq:overlap}
    \mathcal{O}(\boldsymbol{\theta}) = \big|\langle{\psi}|{\varphi}\rangle\big|^2.
\end{equation}
It is worth noting that we measure the performance of the functions $F_H$ and $F_Q$ by the \textit{convergence rate}, i.e., the percentage of problem instances which converged to the values of the overlap greater than or equal to some $\gamma \in [0, 1]$. In our numerical experiments, we set $\gamma = 0.999$.

\begin{figure*}
    \centering
    \includegraphics[width=.325\textwidth]{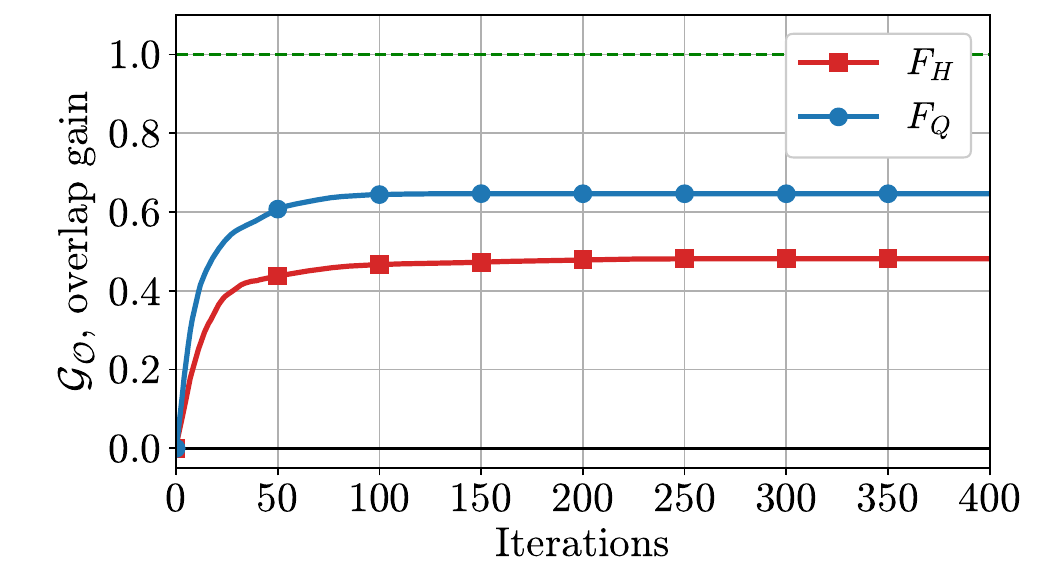}
    \includegraphics[width=.325\textwidth]{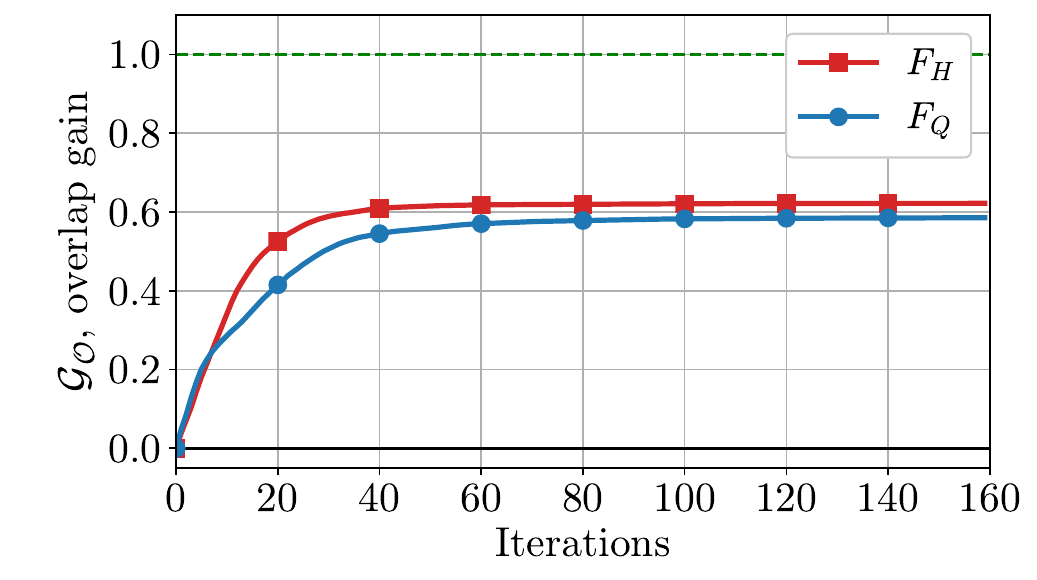}
    \includegraphics[width=.325\textwidth]{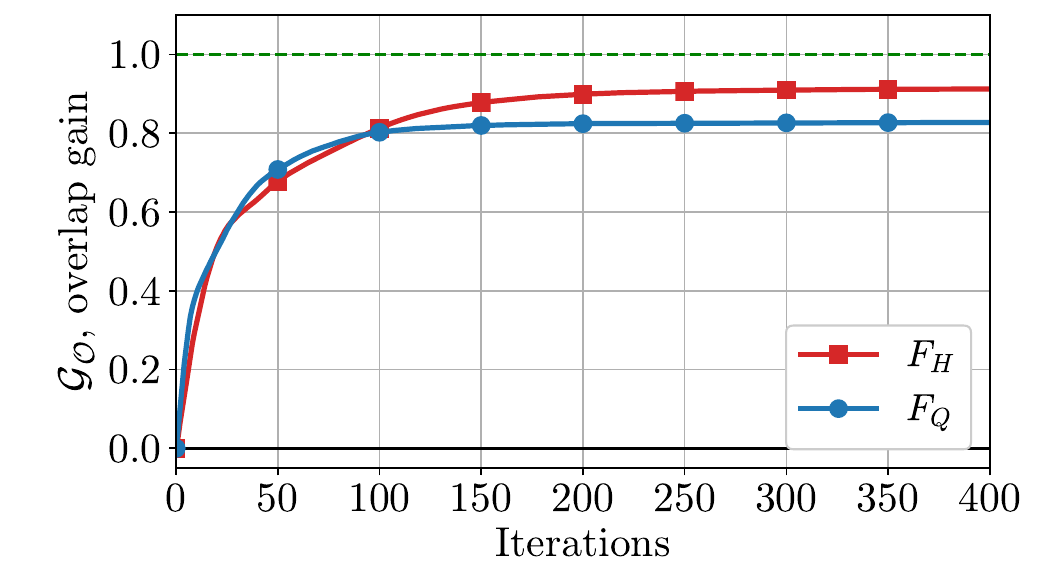}
    \includegraphics[width=.325\textwidth]{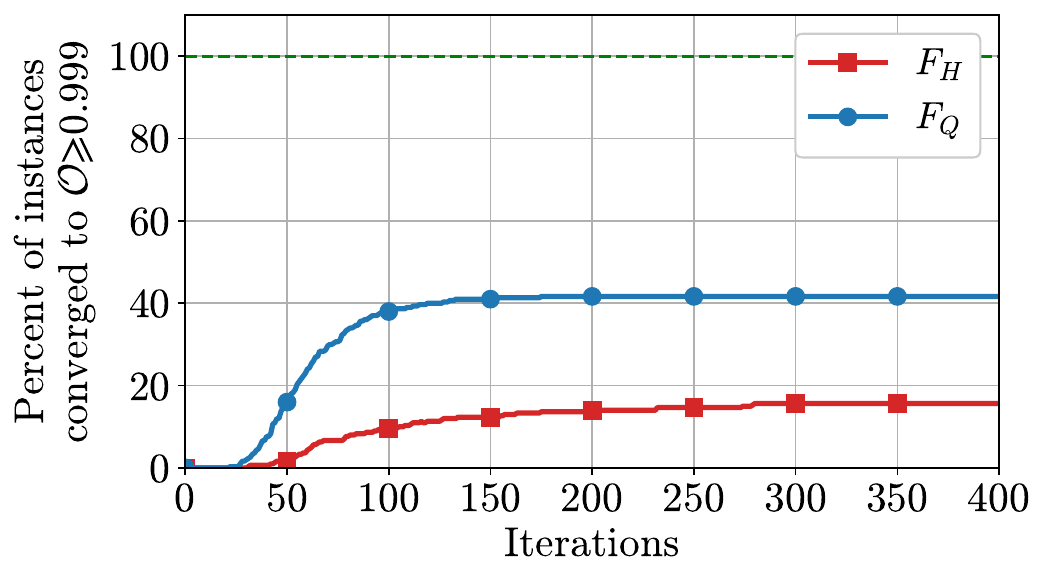}
    \includegraphics[width=.325\textwidth]{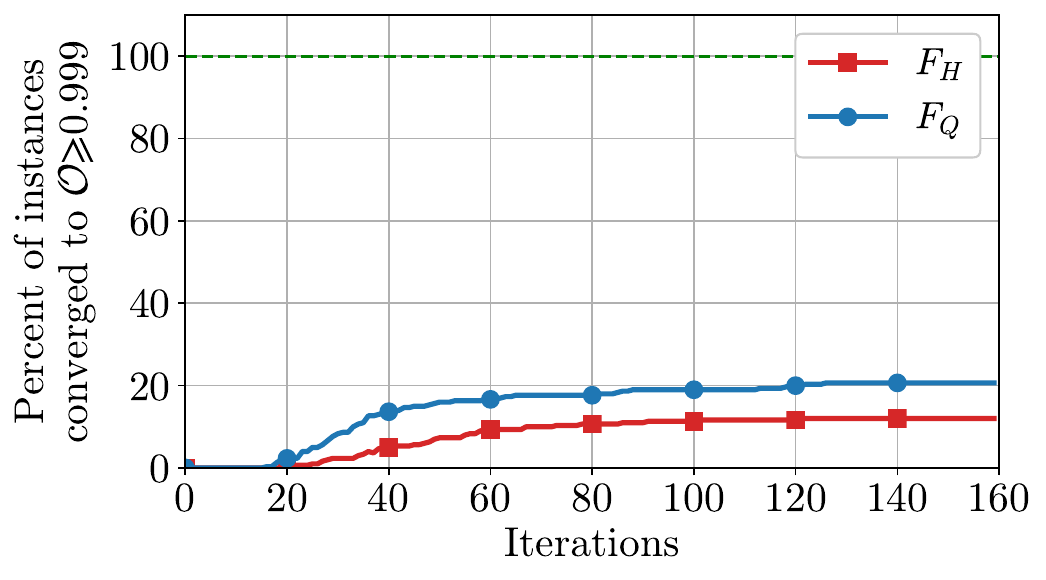}
    \includegraphics[width=.325\textwidth]{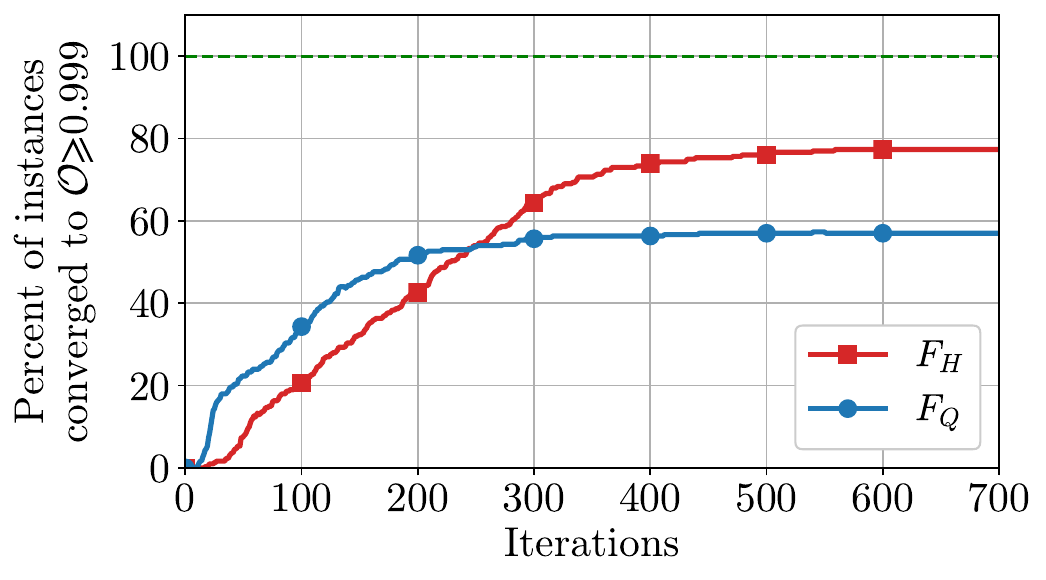}
    \caption{The gain of the overlap between the variational eigenvector ($\ket{\psi}$) and the exact eigenstate ($\ket{\varphi}$) of transverse field Ising model (a), one-dimensional chain of spinless fermions with competing interactions (b), and the massive Schwinger model (c). The corresponding convergence rates are shown in panels (d), (e), and (f), respectively. For each Hamiltonian, 300 random sets of initial parameters for the ansatz were generated.}
    \label{fig:phys_hamiltonians}
\end{figure*}

The plots of the overlap gains and convergence rates for the Hamiltonians of Eqs.~(\ref{eq:tfim})--(\ref{eq:el}) are shown in Fig.~\ref{fig:phys_hamiltonians}.

\begin{figure*}[t]
    \centering
    \includegraphics[width=0.85\textwidth]{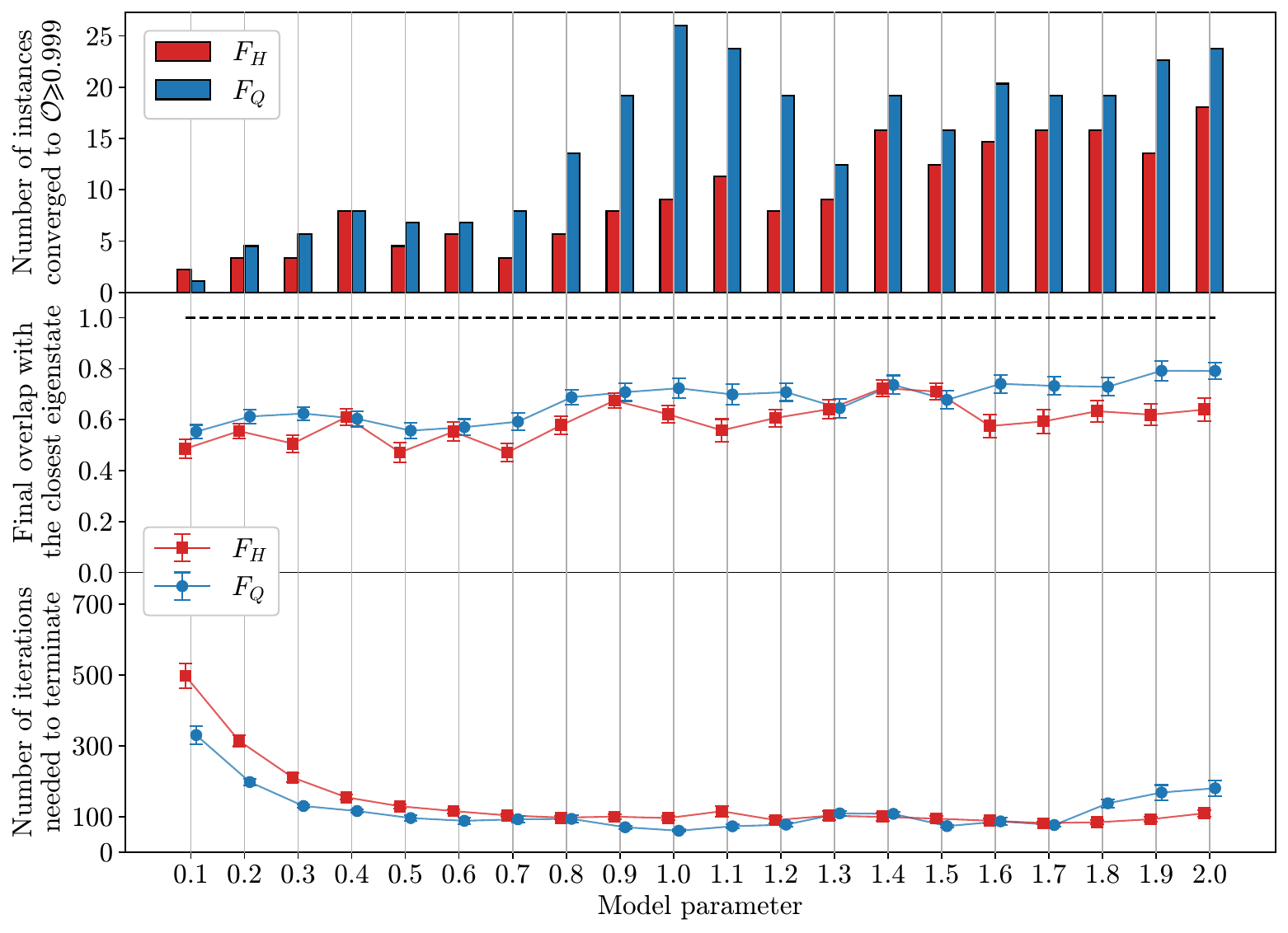}
    \caption{Number of instances (out of 50) that converge to $\mathcal{O} \geqslant 0.999$ (upper panel), overlap with the closest eigenstate (middle panel) after convergence, and the number of iterations needed for the BFGS optimizer to terminate (lower panel) after minimizing the functions $F_H$ (red) and $F_Q$ (blue) for the Ising Hamiltonian. In the middle and lower plots, the solid lines connect mean values of the corresponding functions obtained by averaging more than 50 different realizations of trial states with randomly generated initial parameters as a function of magnetic field (model parameter, $h$) and equipped with standard error bar.}
    \label{fig:ising_performance}
\end{figure*}

As can be visually confirmed, the solution converges suboptimally for both cost functions, but $F_Q$ has a bit higher efficiency in finding the eigenstates of two of the considered Hamiltonians. Note that $F_Q$ is more suitable in dealing with the physical Hamiltonians, specified by Eqs.~(\ref{eq:tfim}) and (\ref{eq:el}), at criticality. Particularly, this is justified by addressing the dependence of optimization performance on the value of the parameter $h$ in the Ising Hamiltonian. It is clearly visible in Fig.~\ref{fig:ising_performance} that both functions $F_H$ and $F_Q$ begin to perform better after $h=0.7$ corresponding to highly correlated state(s). However, at $h=1.1$ the efficiency of $F_H$ drops significantly. On the other hand, the function $F_Q$ exhibits a decreased performance for $h = 1.2, 1.3, 1.4$.

The authors of Ref.~\cite{sung2020exploration} showed the importance of tuning the hyperparameters of different optimizers applied for solving various problems. Since for the objective function \eqref{eq:cost_uni} we can control the evolution time $t$, we could use it as a hyperparameter, making the idea of using $t$ in such a way for the function $F_Q$ to outperform the function $F_H$ in terms of overlap gain or convergence rate viable. For certain types  of Hamiltonians discussed above, there exists $t$ which gives the best performance for $F_Q$. However, as discussed in Appendix~\ref{app:appb}, we did not find any considerable benefit from adjusting the evolution time.

\section{Discussion and conclusion}
The two methods based on minimizing objective functions (\ref{eq:cost_var}) and (\ref{eq:cost_uni}) have substantially different resource costs, as explained in Appendix~\ref{app:appc}. To estimate the variance, one has to know both $\langle\mathcal{H}\rangle$ and $\langle\mathcal{H}^2 \rangle$. Evaluating $\langle\mathcal{H}\rangle$ on a quantum processor requires decomposing a given Hamiltonian $\mathcal{H}$ into the sum of Pauli strings,
\begin{equation}
    \label{eq:hamiltonian_decomposition}
    \mathcal{H} = \sum \mathcal{J}^{ij \dots k}_{\alpha \beta \dots \gamma} \,\sigma^{i}_{\alpha} \sigma^{j}_{\beta} \dots \sigma^{k}_{\gamma},
\end{equation}
and calculating the expectation value of each term separately. In Eq.~(\ref{eq:hamiltonian_decomposition}), upper indices of the real-valued tensor $\mathcal{J}$ denote the qubit number, while the lower indices stand for a specific Pauli operator $\sigma \in \{\mathbb{1}, X, Y, Z\}$. Let us then assume that we need $m$ measurements per Pauli string to achieve predetermined accuracy. If $\mathcal{H}$ contains $k$ Pauli terms, then $\mathcal{H}^2$ contains $k^2$ terms at worst. Thus, we need to run the preparation and measurement circuit about $m(k^2 + k)$ times. This number may be decreased by a smart choice of measurements provided commuting Pauli strings are evaluated simultaneously \cite{verteletskyi_measurement_2019,Yen_Verteletskyi_Izmaylov_2019}. For the needs of quantum chemistry, this approach reduces the number of measurements by an order of $n$, the number of qubits. The number $m$ also has to scale with the number of terms. For one term, the error scales with $1 / \sqrt{m}$, so that $k$ terms would add up to $k / \sqrt{m}$. Thus, to keep the error value fixed, $m$ must scale with the number of Pauli strings, making the number of measurements to be of the order of $k^4 / n$. If we assume that $k$ is at least linear with $n$, the total number of measurements scales as $\mathcal{O}(n^3)$ versus the number of Pauli strings.

The second method applied to $\mathcal{Q}$, on the other hand, requires performing only one set of measurements. The downside is that the quantum circuit is at least twice as deep as that for the variance estimator. On top of that, one needs to be able to implement the Hamiltonian evolution. In the gate model of quantum computation, this can be done using the Suzuki--Trotter formula, which introduces its own error. This means that this technique should require fewer measurements but also a higher degree of gate fidelity. We also note that in order to apply the first method in VQE one has to know the decomposition \eqref{eq:hamiltonian_decomposition} of the target Hamiltonian. At the same time, the second method offers greater utility in the sense that the unitary $\mathcal{Q}$ can be given as a black box quantum circuit, that is, a specific problem whose complexity in terms of gates is not under study.
    
Finally, using these criteria as optimization targets on their own can be helpful for VQE as well. If the solution gets close to some eigenstate, but this state is known not to be the ground state, one can minimize the eigenvalue criteria to get close to that state and then exclude that eigenstate from the search by penalizing overlap with it \cite{Higgott_Wang_Brierley_2019}. Like the VQE, the algorithm we proposed is suitable for noisy intermediate-scale quantum processors, as it does not require the use of ancilla qubits. We believe our method can best employ its potential by accompanying the VQE for verifying a solution, as done in Ref.~\cite{Kokail2019} but by controlling the Hamiltonian's energy variance.

To summarize, using hybrid quantum-classical algorithms remains one of the most promising applications of near-term quantum computers. Within such an approach, one executes as much calculations as possible with classical hardware. VQE is one of the most reliable ways of finding the lowest energy eigenstate of a given matrix. It was recently proposed to make use of mean square deviation to quantify the accuracy of the VQE. In the meantime, such an approach seems to be computationally heavy. In this paper, we proposed a way around this with an objective function which is determined by the evolution operator, or more specifically, a one-parameter unitary group, which appears to be an adequate tool to tackle short-range interacting models that do not require spin-to-qubit mapping.

\acknowledgments
A.K. and J.B. acknowledge support from the research project {\it Leading Research Center on Quantum Computing} (Agreement No.~014/20). The work of A.U. and D.Y. was supported by the Russian Foundation for Basic Research Project No.~19-31-90159.

\appendix

\section{Stability of variational solution and impact of spectral gap on convergence}\label{app:appa}
\begin{figure*}[t]
    \centering
    \includegraphics[width=.325\textwidth]{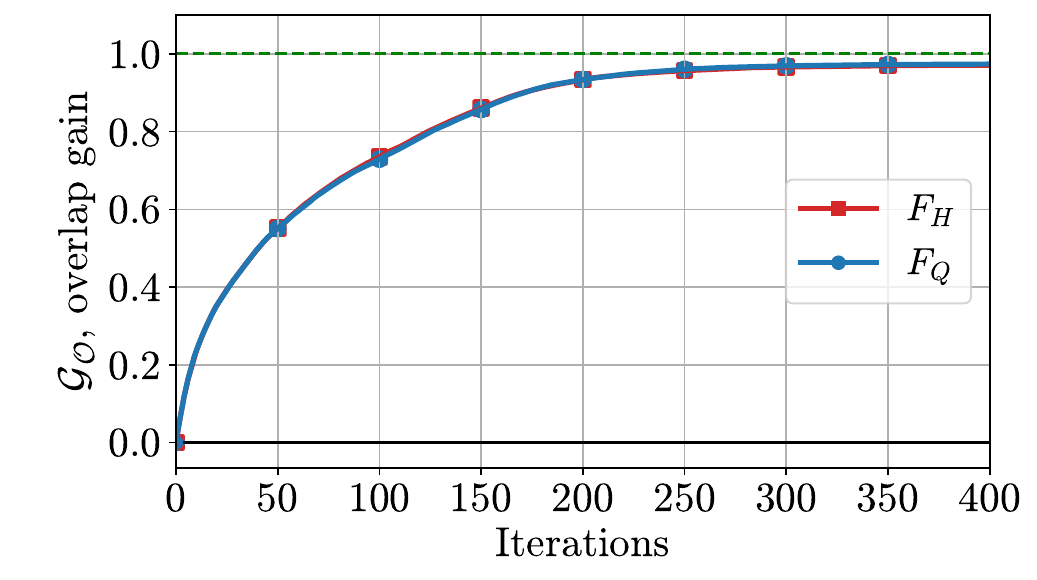}
    \includegraphics[width=.325\textwidth]{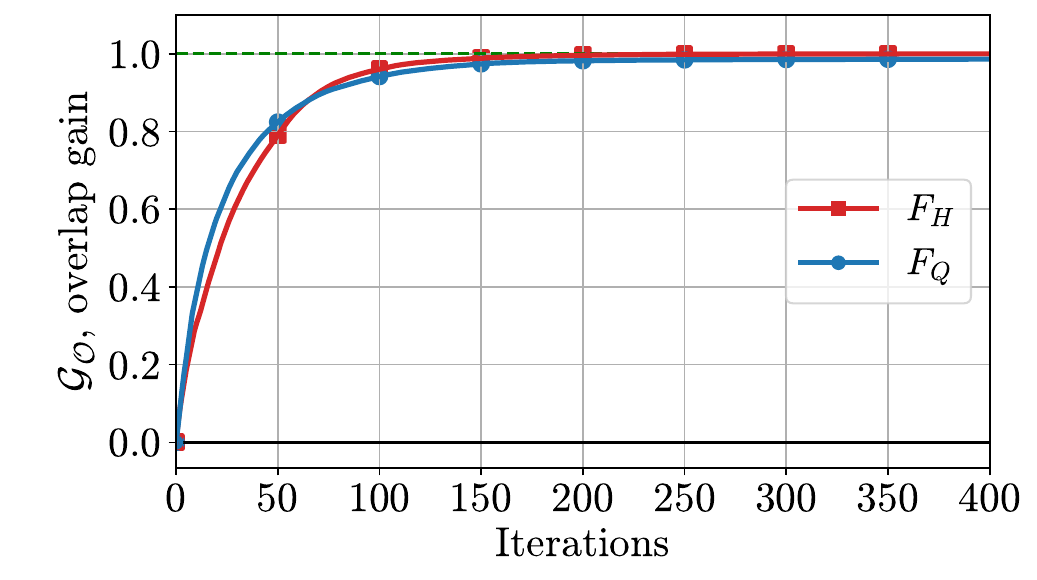}
    \includegraphics[width=.325\textwidth]{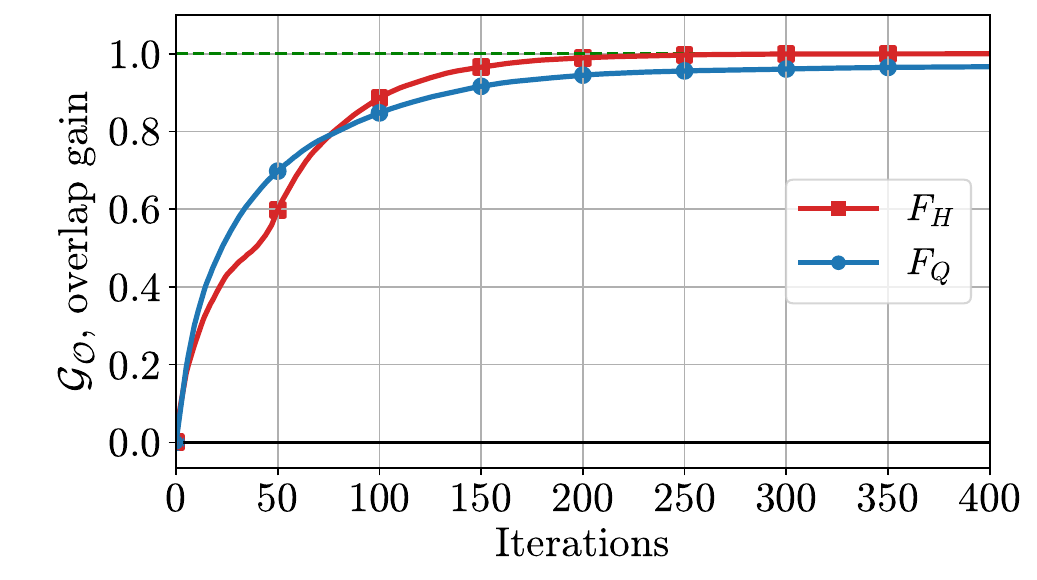}
    \includegraphics[width=.325\textwidth]{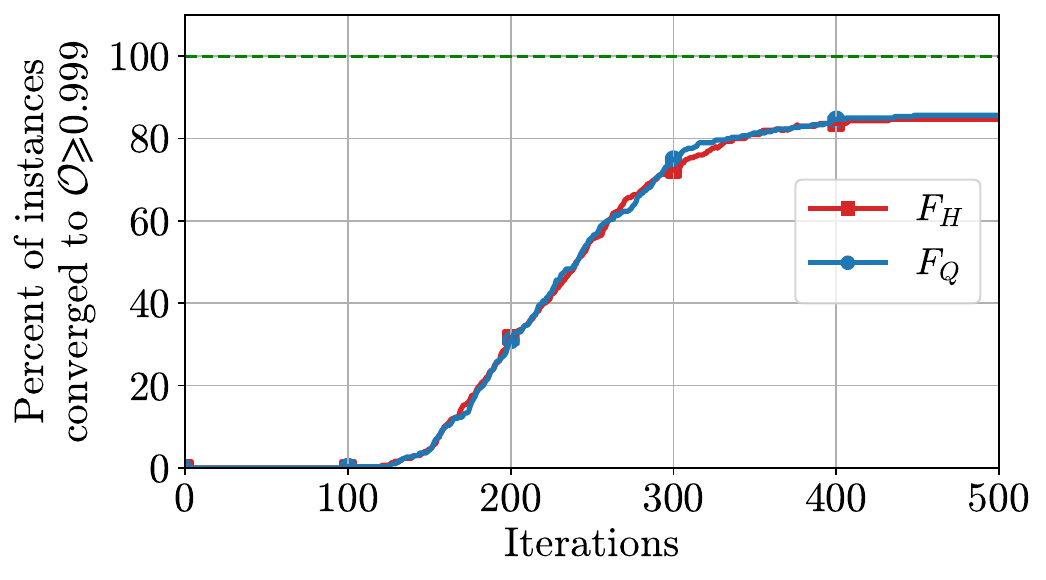}
    \includegraphics[width=.325\textwidth]{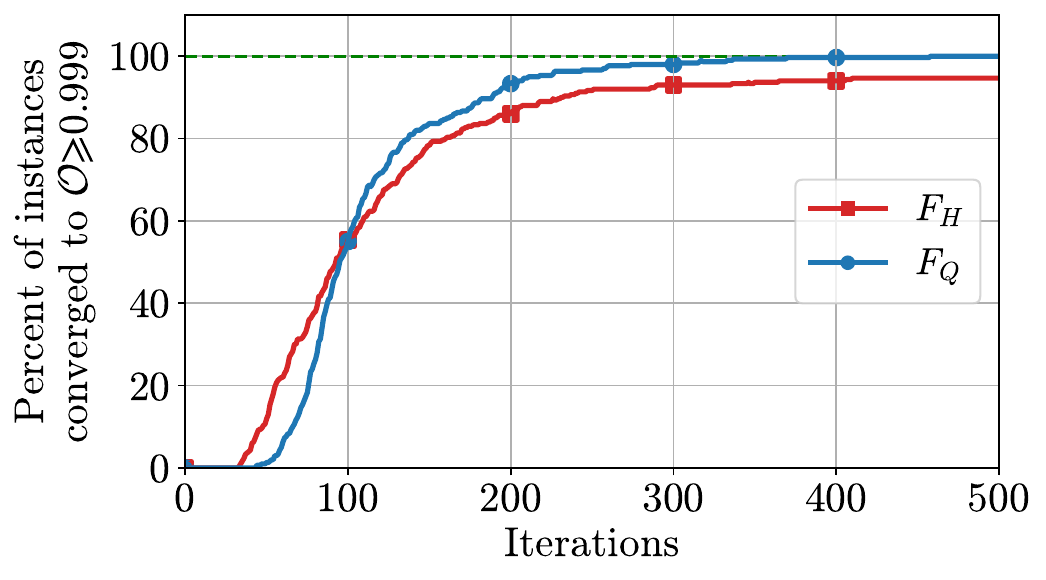}
    \includegraphics[width=.325\textwidth]{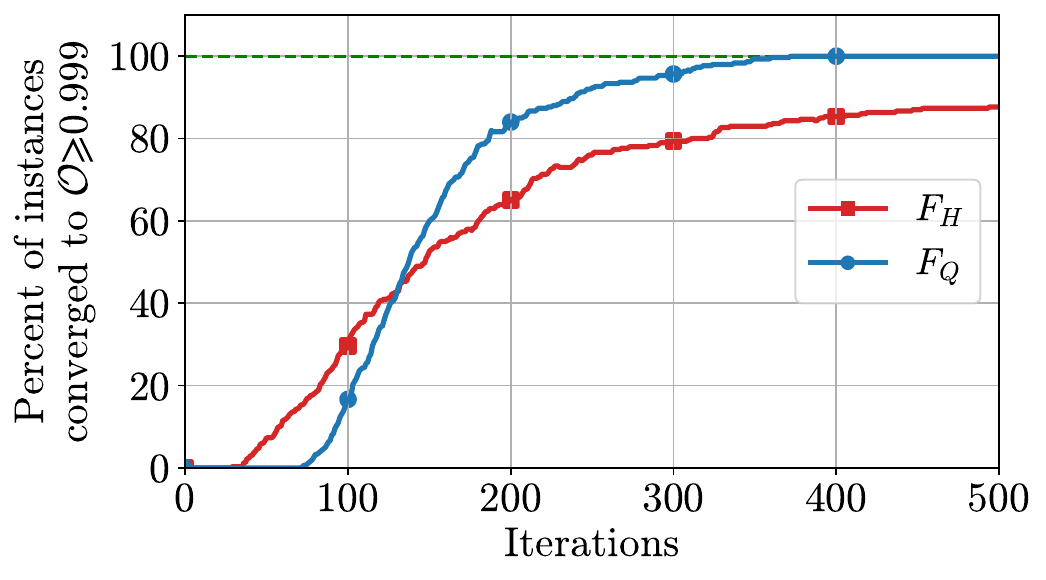}
    \caption{The overlap gains for the multipliers (a) $0.1$, (b) $1$, and (c) $10$ for random Hamiltonians. The corresponding convergence rates are shown in panels (d), (e), and (f), respectively.}
    \label{fig:gap_impact}
\end{figure*}

In the following, we demonstrate that the functions $F_H$ and $F_Q$ have the same performance in dealing with Hamiltonians with small inter-eigenvalue distance. First, we generate 300 random Hamiltonians $\mathcal{H}$ and 300 random sets of initial parameters for the ansatz. We next obtain the convergence rates and the overlap gains for the generated Hamiltonians in three variants: (a) multiplied by $0.1$, (b) the original ones, and (c) multiplied by $10$. The corresponding plots are illustrated in Fig.~\ref{fig:gap_impact}. We note that for the function $F_Q$, multiplying the Hamiltonian by a real number is equivalent to setting the evolution time $t$ since $\mathcal{Q} = e^{i\mathcal{H}t}$. 

In the vicinity of an eigenstate, this problem allows an analytical treatment. In fact, using the spectral theorem for the target Hamiltonian, $\mathcal{H} = \sum_{j=1}^n \lambda_j \ket{\lambda_j}\bra{\lambda_j}$, we can rewrite the functions \eqref{eq:cost_var} and \eqref{eq:cost_uni} as follows:
\begin{align*}
    F_H &=  \sum\limits_{ij}\lambda_i(\lambda_i-\lambda_j)|\beta_i|^2|\beta_j|^2, \\
    F_Q &=  2\sum\limits_{ij}|\beta_i|^2|\beta_j|^2\sin^2\left(\frac{(\lambda_i-\lambda_j)t}{2}\right),
\end{align*}
where $\beta_j = \langle\lambda_j|\psi\rangle$. For the sake of simplicity, we assume $0<\lambda_1<\lambda_2<\lambda_3<\ldots<\lambda_n$. Looking at the equations above, one may expect that as the average distance between the eigenvalues of $\mathcal{H}$, $(\lambda_j - \lambda_k)$, decreases, the less the functions differ from each other less. Therefore, provided that the target Hamiltonian has small intereigenvalue distances, the functions $F_H$ and $F_Q$ show the same efficiency in finding an eigenvector. The VQE solution in the neighborhood of the state $\ket{\lambda_1}$ may be written as
\begin{equation}
\label{eq:variational_state_decomposition}
\ket{\psi}=\frac{1}{\sqrt{1+\epsilon^2}}\ket{\lambda_1}+ \frac{\epsilon}{\sqrt{1+\epsilon^2}} \ket{\phi},
\end{equation}
with $\epsilon\ll1$. The vector $\ket{\phi}=\sum_{j>1} c_j\ket{\lambda_j}$ is normalized, so the squares of absolute values of $c_j$ sum to unity. Consider the variance $\delta=F_H(\psi)$ as given by Eq.~\eqref{eq:cost_var},

\begin{equation}\label{eqn:eps-variance} 
\delta = (\lambda_1^2 + d^2 - 2 e\lambda_1)\epsilon^2
- (d^2 + e^2 + 2 \lambda_1^2 - 4 e\lambda_1)\epsilon^4 + o(\epsilon^4),
\end{equation}
on the condition that $d^2=\bra{\varphi}\mathcal{H}^2 \ket{\varphi}=\sum_{j>1} |c_j|^2 \lambda_j^2$ and $e=\bra{\varphi}\mathcal{H}\ket{\varphi}=\sum_{j>1} |c_j|^2 \lambda_j$. Conversely, one can treat Eq.~(\ref{eqn:eps-variance}) as an implicit function $\epsilon = \epsilon (\delta)$. By considering the derivatives of this function in the vicinity of $\epsilon = 0, \delta = 0$, we arrive at
\begin{equation}
    \epsilon^2 = \frac{\delta}{\lambda_1^2 + d^2 - 2 e\lambda_1} \\ + \frac{3(d^2+e^2+2\lambda_1^2-4e\lambda_1)}{(\lambda_1^2+d^2-2e\lambda_1)^3}\delta^2+ o(\delta^2),
\end{equation}
or, alternatively,
\begin{equation}
    \delta = \epsilon^2 \sum_{j>1} |c_j|^2 (\lambda_j - \lambda_1)^2 + o(\epsilon^2).
\end{equation}
Suppose that we search for an eigenvector of a unitary $Q = e^{-i\mathcal{H}t}$ for some Hermitian $\mathcal{H}$ and real $t$, and we can implement this $\mathcal{H}$ evolution. Assume that $\ket{\psi}$ is sufficiently close to an eigenvector, and $\big|\bra{\psi} e^{-i\mathcal{H}t} \ket{\psi}\big| = 1 - \tilde{\delta}$, where 
\begin{equation}
    \tilde{\delta} = 2 \epsilon^2 \Big[1 
    - \sum_{j>1} |c_j|^2 \cos\big[(\lambda_j - \lambda_1)t\big]
    \Big]  + o(\epsilon^2). 
\end{equation}
Notice that $\tilde{\delta}$ is a function of time $t$. Using this dependence, we can extract some extra properties of the target Hamiltonian.

\begin{figure*}
    \centering
    \includegraphics[width=.65\textwidth]{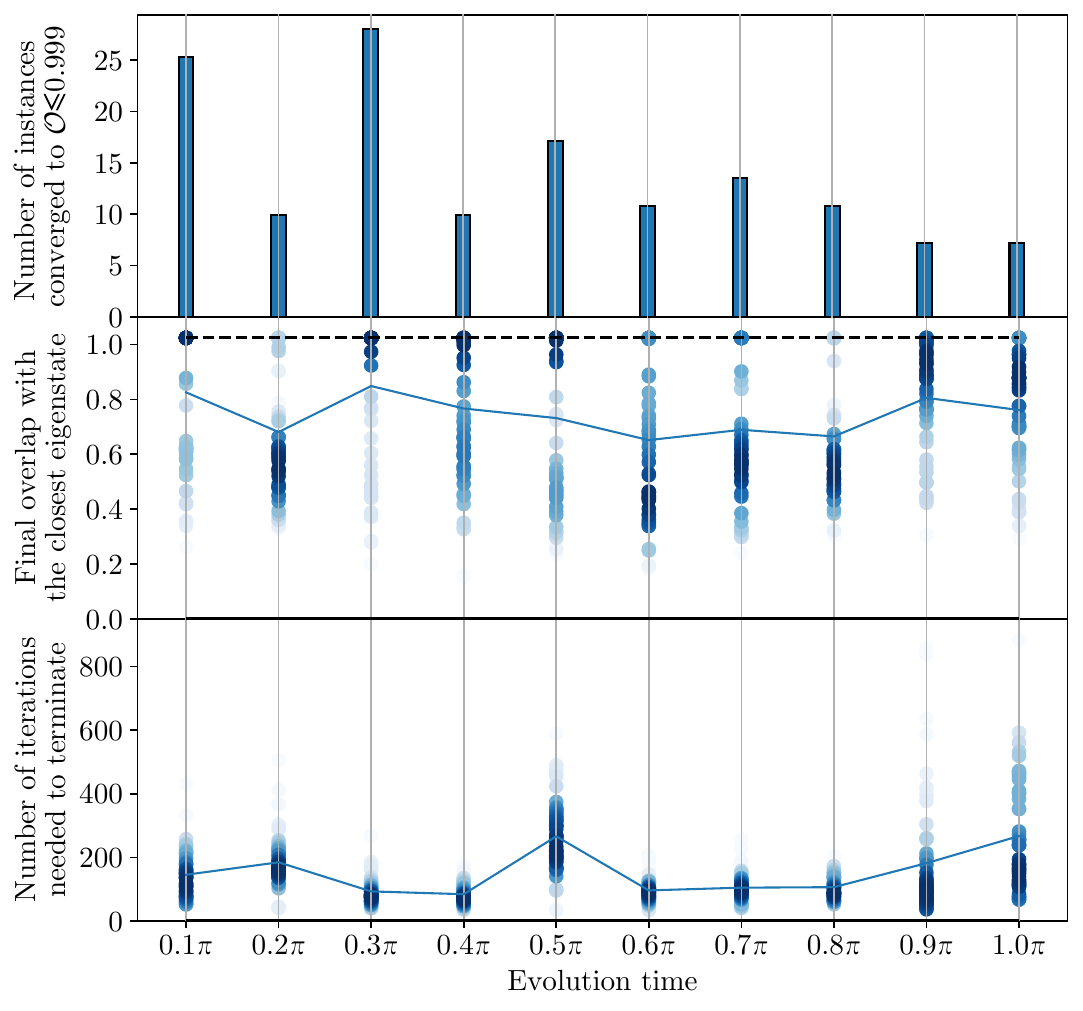}
    \caption{Number of instances (out of 50) converged to $\mathcal{O} \geqslant 0.999$ (upper), final overlap with the closest eigenstate (middle), and the number of iterations needed for the BFGS optimizer to terminate (lower) after minimizing the function $F_Q$ for the Ising Hamiltonian. In the middle and lower plots, for each evolution time, there are 50 data points, one for each set of randomly generated initial parameters for the ansatz. The color intensity of the circles corresponds to the density of data points. The solid line connects the mean values.}
    \label{fig:evolution_time}
\end{figure*}

\section{Evolution time as a hyperparameter}\label{app:appb}
In the main text, we emphasized that there could be an optimal evolution time parameter to be used in $F_Q$ for specific problem instances. To give quantitative arguments, we provide BFGS optimization performance versus the evolution time in Fig.~\ref{fig:evolution_time}. To illustrate our findings, we consider the Hamiltonian of transverse field Ising model given by Eq.~(\ref{eq:tfim}) at criticality $h=1$,
\begin{equation*}
    \mathcal{H}_{\text{TFIM}} = \sum_j \left( Z_j Z_{j+1} + X_j \right),
\end{equation*}
as the target. Our numerical findings do not support the idea that any significant advantage can be achieved by tuning the evolution time. However, some values of $t$, e.g., $0.1\pi$, $0.5\pi$, or $0.9\pi$, allow us to get a slightly better performance. We also note that the best results are obtained for $t=0.3\pi \approx 1$.

\section{Costs for implementing minimization}\label{app:appc}

Here we analyze costs needed for evaluating the functions $F_H$ and $F_Q$ on a noisy intermediate-scale quantum hardware in terms of circuits and gates. As an example, we consider the transverse field Ising model of $n$ spins as given by Eq.~\eqref{eq:tfim} in the main text. One can relatively easy develop the unitary evolution using the $r$-step first-order Trotter decomposition:
\begin{equation}
    \label{eq:trotter}
    U\equiv e^{i \mathcal{H}_{\text{TFIM}} t } \approx
    \left( \prod_{j=1}^n e^{i \mathcal{J} t Z_j Z_{j+1} /r}  \prod_{j=1}^n e^{i \mathcal{J} h t X_j /r} \right)^r
\end{equation}
with $Z_n Z_{n+1} \equiv Z_1 Z_n$. Provided that one can implement $ZZ$ rotations on a given quantum device, the circuit construction for $U$ is straightforward. If this is the case, this circuit is constituted by $N_U=2nr$ gates, $nr$ of which correspond to $ZZ$ rotations and the others to $X$ rotations. 
    
Suppose we use an $l$-layered hardware-efficient ansatz $V$ with $N_V = 4ln$ gates. Measuring $F_Q = | \langle V^\dagger U V \rangle |^2$ requires $N_Q = 2N_V + N_U = 2n(4l + r)$ gates, i.e., scales linearly with $n$. In contrast, to evaluate $\langle \mathcal{H} \rangle$ one has to have $2n$ circuits---one for each term in the target Hamiltonian (\ref{eq:tfim}). Each circuit consists of $N_V$ gates of the ansatz, and each second circuit possesses an additional Hadamard gate for measuring the $X_j$ terms. Overall, for $n > 4$, one needs $N_{H_1} = 2n^2(4l + 1)$ gates which is quadratic in $n$. On the other hand, $\mathcal{H}^2$ has $(2n^2 - 3n + 1)$ terms, requiring thus this number of circuits to be implemented---each circuit contains $N_V$ gates coming from the ansatz as well as $(2n^2 - 3n)$ additional Hadamard gates for measuring each $X$ operator. This, for $n>5$, results in $N_{H_2} = (-3 + 4 l) n + (2 - 12 l) n^2 + 8 l n^3$ gates for measuring $\langle \mathcal{H}^2 \rangle$. Overall, one needs
\begin{equation*}
    N_H = N_{H_1} + N_{H_2} = (4l - 2) (n - n^2) + 8 l n^3
\end{equation*}
gates for calculating $F_H$.
    
Comparing $N_Q$ and $N_H$, one can clearly deduce that using $F_Q$ as a cost function is superior to $F_H$ in terms of the total number of gates as long as $r$ scales as $\mathcal{O}(n^\alpha)$, where $\alpha<2$, with the number of qubits $n$. Note, however, that $N_H$ gates are ``distributed'' among $(2n^2 - n + 1)$ circuits, whereas all the $N_Q$ gates are composed into one circuit which may potentially cause a lower performance for simulating $F_Q$ function on a noisy quantum hardware. 

\begin{figure*}
    \centering
    \includegraphics[width=.695\textwidth]{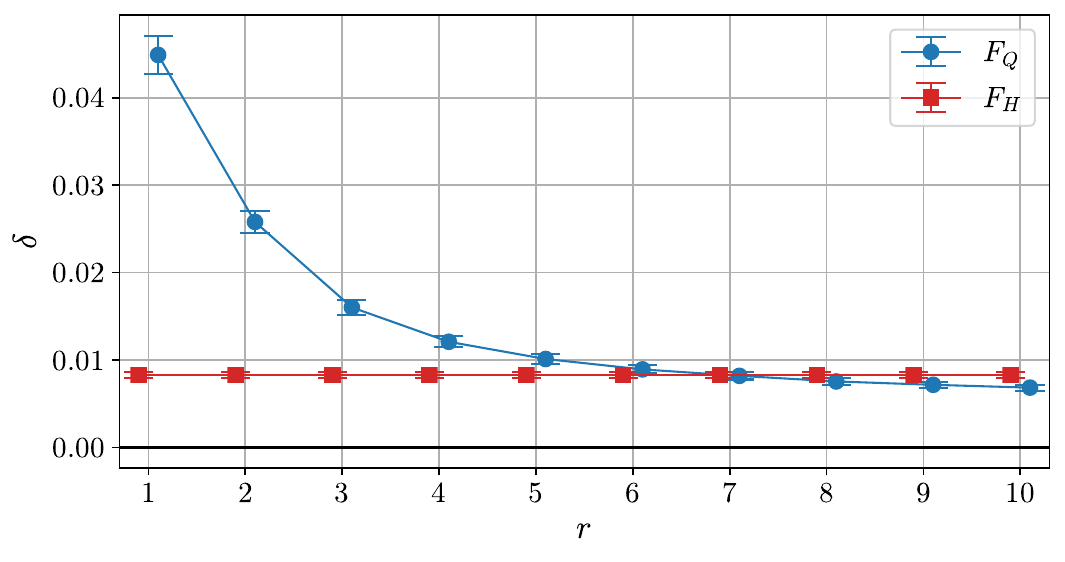}
    \caption{The absolute difference between the approximate and exact values of objective functions, $\delta$, vs the number of Trotter steps $r$. The value for $F_H$ is normalized, while each data point is obtained after averaging over $300$ instances. The error bars indicate the standard errors.}
    \label{fig:delta-k}
\end{figure*}

To analyze the error gained during the calculation of the objective functions we define approximate $F_{H/Q}^\mathrm{app}$, which are estimated using the {\tt Qiskit} package \cite{Qiskit}, and exact $F_{H/Q}^\mathrm{ex}$ values of the cost functions. Note that {\tt Qiskit} allows one to emulate the finite number of measurements $m$ performed for each circuit---for the purposes of our simulations, we set this number $m=1000$. $F_{H/Q}^\mathrm{ex}$ are obtained without imitating finite statistics, in other words as if we let $m \rightarrow \infty$, and with no Trotter decomposition implemented for $F_Q$. To provide a quantitative estimate, we plot the absolute difference $\delta_{H/Q}(r)=|F_{H/Q}^\mathrm{ex} - F_{H/Q}^\mathrm{app}|$ depending on the number of repetitions in \eqref{eq:trotter} for a five-qubit TFIM Hamiltonian at criticality on condition a four-layered hardware-efficient ansatz is used; see Fig.~\ref{fig:delta-k}. One can clearly notice that despite ``{\it trotterization}'' even for $r=8$ $\delta_Q(r)$ lowers down as compared to $\delta_H(r)$. Moreover, calculating $F_H^\mathrm{app}$ requires for $46$ circuits to be evaluated---each of which contains $80$ gates from the ansatz and some number of the Hadamard gates, i.e., $N_H = 3720$ gates in total. In contrast, one has to have only one circuit with $N_Q = 260$ gates for calculating $F_Q^\mathrm{app}$ with $r=10$.
    
However, one has to be aware of the fact that this could potentially be not the case for Hamiltonians with high degree of nonlocality which agrees well with recent findings \cite{commeau2020variational}. For example, the Hamiltonian of $n$ spinless fermions with competing interactions, as given by Eq.~(\ref{eq:el}) in the main text, is characterized by the presence of nonlocal terms (e.g., $Z^{\otimes(n-1)} \otimes \sigma^\pm$) after spin-to-qubit mapping being done. Furthermore, it would be hard to decompose the unitary evolution of this term into a sequence of two-qubit gates on a real piece of quantum hardware.

\bibliographystyle{apsrev4-2}

\end{document}